\documentclass[10pt,twocolumn,showpacs,amsmath,amssymb,floatfix,superscriptaddress]{revtex4-2}

\usepackage{graphicx}
\usepackage{dcolumn}
\usepackage{bm}
\usepackage{color}
\usepackage{txfonts}
\usepackage{microtype}
\usepackage{hyperref}
\usepackage[english]{babel}
\usepackage{slashed}
\usepackage{gensymb}
\usepackage{epsfig}
\usepackage[normalem]{ulem}
\usepackage{amsmath}
\usepackage{array}
\usepackage{booktabs}
\usepackage{multirow}
\allowdisplaybreaks[4]

\begin{document}
\renewcommand{\figurename}{FIG}	

\title{Manipulation of $\gamma$-ray polarization in Compton scattering}

\author{Yu Wang}
\affiliation{Ministry of Education Key Laboratory for Nonequilibrium Synthesis and Modulation of Condensed Matter, Shaanxi Province Key Laboratory of Quantum Information and Quantum Optoelectronic Devices, School of Physics, Xi'an Jiaotong University, Xi'an 710049, China}
\author{Mamutjan Ababekri}
\affiliation{Ministry of Education Key Laboratory for Nonequilibrium Synthesis and Modulation of Condensed Matter, Shaanxi Province Key Laboratory of Quantum Information and Quantum Optoelectronic Devices, School of Physics, Xi'an Jiaotong University, Xi'an 710049, China}
\author{Feng Wan}
\affiliation{Ministry of Education Key Laboratory for Nonequilibrium Synthesis and Modulation of Condensed Matter, Shaanxi Province Key Laboratory of Quantum Information and Quantum Optoelectronic Devices, School of Physics, Xi'an Jiaotong University, Xi'an 710049, China}
\author{Jia-Xing Wen}
\affiliation{Key laboratory of plasma physics, Research center of laser fusion, China academy of engineering physics, Mianshan Rd 64\#, Mianyang, Sichuan 621900, China}
\author{Wen-Qing Wei}
\affiliation{Ministry of Education Key Laboratory for Nonequilibrium Synthesis and Modulation of Condensed Matter, Shaanxi Province Key Laboratory of Quantum Information and Quantum Optoelectronic Devices, School of Physics, Xi'an Jiaotong University, Xi'an 710049, China}
\author{Zhong-Peng Li}
\affiliation{Ministry of Education Key Laboratory for Nonequilibrium Synthesis and Modulation of Condensed Matter, Shaanxi Province Key Laboratory of Quantum Information and Quantum Optoelectronic Devices, School of Physics, Xi'an Jiaotong University, Xi'an 710049, China}
\author{Hai-Tao Kang}
\affiliation{Ministry of Education Key Laboratory for Nonequilibrium Synthesis and Modulation of Condensed Matter, Shaanxi Province Key Laboratory of Quantum Information and Quantum Optoelectronic Devices, School of Physics, Xi'an Jiaotong University, Xi'an 710049, China}
\author{Bo Zhang}\email{zhangbolfrc@caep.cn}
\affiliation{Key laboratory of plasma physics, Research center of laser fusion, China academy of engineering physics, Mianshan Rd 64\#, Mianyang, Sichuan 621900, China}
\author{Yong-Tao Zhao}
\affiliation{Ministry of Education Key Laboratory for Nonequilibrium Synthesis and Modulation of Condensed Matter, Shaanxi Province Key Laboratory of Quantum Information and Quantum Optoelectronic Devices, School of Physics, Xi'an Jiaotong University, Xi'an 710049, China}
\author{Wei-Min Zhou}\email{zhouwm@caep.cn}
\affiliation{Key laboratory of plasma physics, Research center of laser fusion, China academy of engineering physics, Mianshan Rd 64\#, Mianyang, Sichuan 621900, China}
\author{Jian-Xing Li}\email{jianxing@xjtu.edu.cn}
\affiliation{Ministry of Education Key Laboratory for Nonequilibrium Synthesis and Modulation of Condensed Matter, Shaanxi Province Key Laboratory of Quantum Information and Quantum Optoelectronic Devices, School of Physics, Xi'an Jiaotong University, Xi'an 710049, China}

\date{\today}
	
\begin{abstract} 
High-brilliance high-polarization $\gamma$ rays based on Compton scattering are of great significance in broad areas, such as nuclear, high-energy, astro-physics, etc. However, the transfer mechanism of spin angular momentum in the transition from linear, through weakly into strongly nonlinear processes is still unclear, which severely limits the simultaneous control of brilliance and polarization of high-energy $\gamma$ rays. In this work, we investigate the manipulation mechanism of high-quality polarized $\gamma$ rays in Compton scattering of the ultrarelativistic electron beam colliding with an intense laser pulse. We find that the contradiction lies in the simultaneous achievement of high-brilliance and high-polarization of $\gamma$ rays by increasing laser intensity, since the polarization is predominately contributed by the electron (laser photon) spin via multi-photon (single-photon) absorption channel. Moreover, we confirm that the signature of $\gamma$-ray polarization can be applied for observing the nonlinear effects (multi-photon absorption) of Compton scattering with moderate-intensity laser facilities.
\end{abstract}
	
\maketitle
Polarized $\gamma$ rays are powerful probes for basic researches and applications \cite{Simon1950,FAGG1959,Zernike1955,Rousse2001,Dean2008}. For instances, polarized photons below 1 MeV enable the crystallographic probing and the biomedical imaging with femtosecond time resolution \cite{Rousse2001}. In the energy range of several MeV to tens of MeV, polarized photo-induced nuclear reactions are highly effective in studying nuclear physics, transmutation and astrophysics \cite{Schwoerer2006,Weller2009,Budker2022,Zilges2022}. Furthermore, polarized $\gamma$ rays can be readily deployed for testing the interaction between two real photons leading to the linear Breit-Wheeler (BW) electron-positron pair production \cite{Breit1934,Zhao2022,Zhao2023} and photon-photon elastic scattering \cite{Halpern1933,Micieli2016}. For energy scales ranging from hundreds of MeV to GeV, polarized $\gamma$ rays are significant for probing vacuum birefringence \cite{Wistisen2013,Bragin2017,Wan2022}.

Traditionally, polarized $\gamma$-ray sources are mainly obtained via synchrotron radiation \cite{Rousse2001,Bilderback2005}, bremsstrahlung \cite{Schwengner2005, Sonnabend2011}, and linear Compton scattering (LCS) \cite{Kawase2008,Amano2009,Ur2015}. For synchrotron radiation facilities, as the energy of $\gamma$ rays $\varepsilon_\gamma \propto \varepsilon_e^2/\lambda$ \cite{Bech2009,Als2011}, higher-energy electrons are required because of the insertion devices of undulators (wigglers) with the wavelength $\lambda$ of a few centimeters, where $\varepsilon_e$ is the electron energy. Compared with bremsstrahlung, LCS is characterized by good directionality, collimation and high polarization \cite{Albert2010,Sarri2014}. However, the low scattering probability sets a practical limit for the maximum flux no larger than 10$^{10}$ photons$\cdot$s$^{-1}$ \cite{Budker2022}.

Recently, the rapidly advanced high-power laser technique \cite{Gales2018,Danson2019,Yoon2019,Yoon2021} has promoted the research of high-energy high-brilliance polarized $\gamma$ rays \cite{Sarri2014,Khrennikov2015,Yan2017}, where the interaction mechanism transitions from linear into the nonlinear regime \cite{Bula1996,Ivanov2004,Yousef2006,Xiao2023}. At intermediate laser intensity, polarized $\gamma$ rays can be generated by spin-nonpolarized (SNP) electron beams via weakly nonlinear CS (NLCS) \cite{Wistisen2019,King2020,Tang2020,Wang2022}. Furthermore, the stronger laser field scattered with initially spin-polarized (SP) electron beams enables the generation of more brilliant high-polarization $\gamma$ rays in strongly NLCS \cite{Li2020}. Importantly, in the near future, experiments such as LUXE at DESY \cite{Abramowicz2021} and E320 at FACET-\uppercase\expandafter{\romannumeral2} \cite{Meuren2019} will be performed using the conventionally accelerated $\varepsilon_{e} \sim$ 10 GeV electron beam in collision with dozens up to hundreds of TW (corresponding to the laser invariant intensity $a_0 \sim 10$) laser pulse to probe the transition from linear to strongly nonlinear QED regime. Moreover, all-optical PW up to 10 PW laser facilities have also been commissioned or will be online \cite{Bruno2014,Zou2014,Gan2021,Gonoskov2022,Fedotov2023}. However, there is no charted transfer mechanism of spin angular momentum (SAM) in the transition from linear, through weakly into strongly nonlinear CS. Therefore, controlling the brilliance and polarization of high-energy $\gamma$ rays is still a great challenge. 

\begin{figure}[t] 
	\begin{center}
		\includegraphics[width=\linewidth]{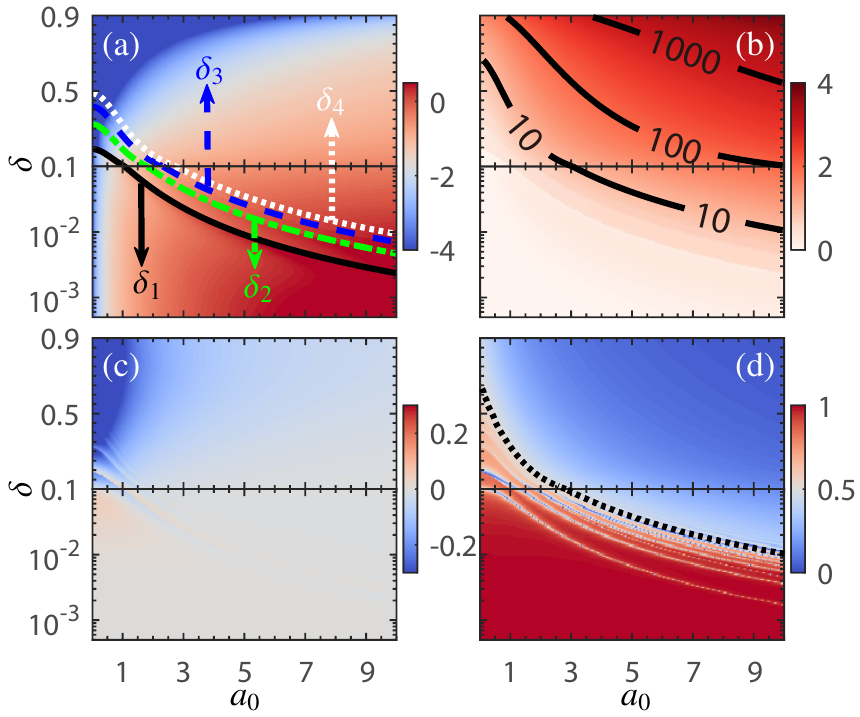}
		\caption{
			(a) Distribution of the total differential transition rate log$_{10} \frac{\mathrm{d^2}W_{rad}}{\mathrm{d}t \mathrm{d}\delta}$ after summing over the photon helicity and averaging over the initial electron spin versus the laser invariant intensity $a_0$ and energy ratio parameter $\delta$, where $\frac{\mathrm{d}^2W_{rad}}{\mathrm{d}t\mathrm{d}\delta} = 2W_0 \sum_{n=1}^{\infty} F_{1n}$ with $\omega_L=1.55$ eV and $\varepsilon_{e}=10$ GeV. The black-solid, green-dash-dotted, blue-dashed, and white-dotted lines indicate the first four cutoffs $\delta_{n=1,2,3,4}$, respectively. (b) Average number of absorbed laser photons log$_{10} \overline{n}$ versus $a_0$ and $\delta$. The black-solid lines indicate the contour lines of $\overline{n}=$ 10, 100 and 1000, respectively. Ratios of (c) $\frac{F_2}{F_1}$ and (d) $\frac{\left|F_3\right|}{\left|F_3\right|+F_4}$ ($F_4 >0$) versus $a_0$ and $\delta$, respectively. The black-dotted line in (d) indicates the spin contribution of the laser is equal to that of an electron to the $\gamma$ photon, i.e., $\frac{\left|F_3\right|}{\left|F_3\right|+F_4}$=0.5.  }\label{fig1}
	\end{center}
\end{figure}

\begin{figure}[t] 
	\begin{center}
		\includegraphics[width=\linewidth]{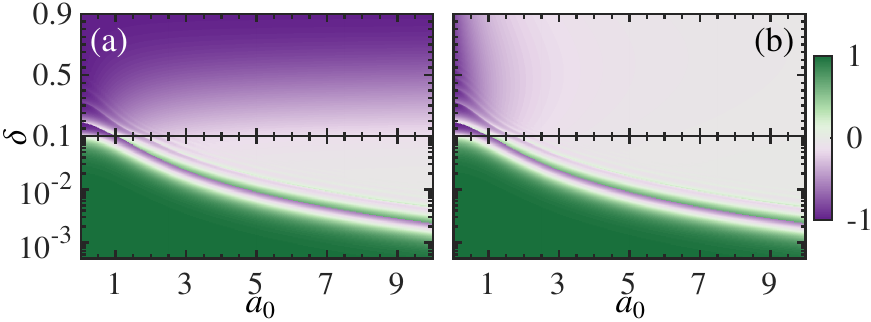}
		\caption{
			Average helicity $\overline{h}_\gamma$ of the emitted photons versus $a_0$ and $\delta$ for two different models of (a) the laser helicity $h_L = 1$ and initial electron helicity $h_e = -1$, and (b) $h_L = 1$ and $h_e = 0$, respectively. The laser and electron-beam parameters are the same as those in Fig.\ref{fig1}.}\label{fig2}
	\end{center}
\end{figure}

\begin{figure}[t] 
	\begin{center}
		\includegraphics[width=\linewidth]{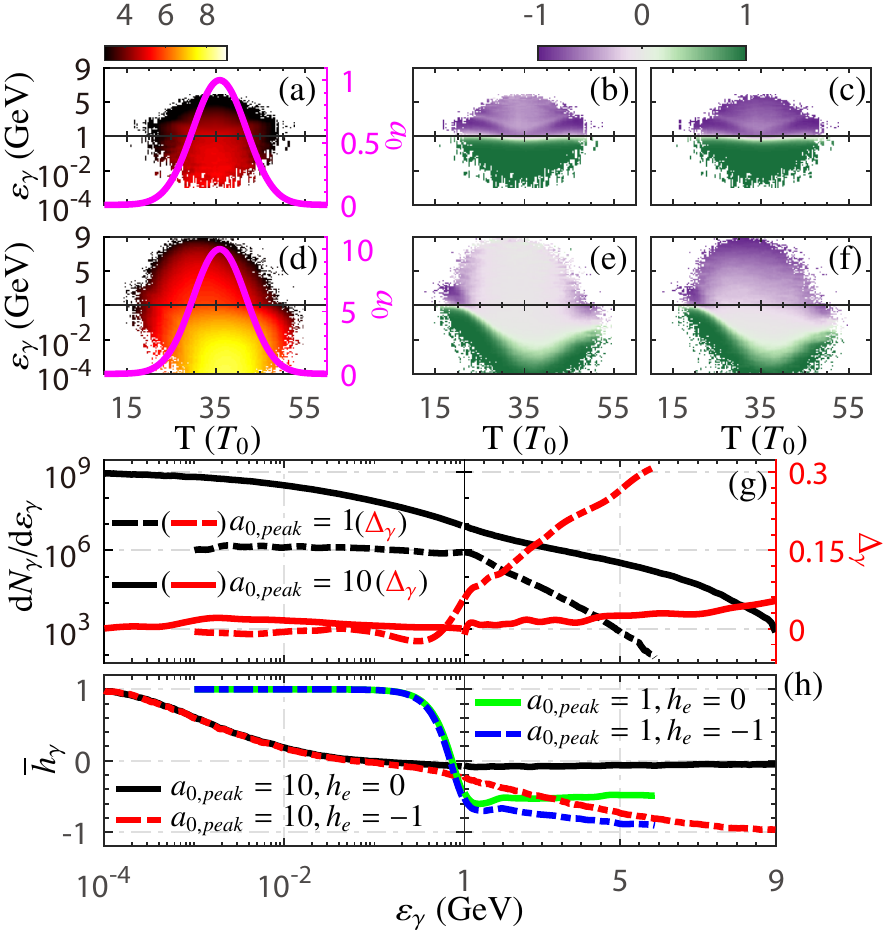}
		\caption{
			(a)-(c) Generation of $\gamma$ photons in the realistic laser pulse with peak intensity $a_{0,peak}=1$ versus the interaction time T ($T_0$) and $\gamma$-photons energy $\varepsilon_{\gamma}$ with distributions of (a) and (b) the yields log$_{10} \frac{\mathrm{d}N_\gamma}{\mathrm{dTd}\varepsilon_{\gamma}}$ and average helicity $\overline{h}_\gamma$ for $h_e=0$, and (c) $\overline{h}_\gamma$ for $h_e=-1$, respectively, where $T_0$ is the laser period. The magenta solid line in (a) indicates the laser invariant intensity $a_0$ in real-time versus T ($T_0$). (d)-(f) The physical representations and other laser and electron-beam parameters are the same as those in (a)-(c) respectively, except $a_{0,peak}=10$. Comparisons of (g) energy spectra d$N_\gamma$/d$\varepsilon_{\gamma}$ with relative deviation $\Delta_{\gamma}$ and (h) $\overline{h}_\gamma$ between the cases of $a_{0,peak}=1$ and $a_{0,peak}=10$ versus $\varepsilon_{\gamma}$, respectively, where $\Delta_{\gamma}=(\mathcal{N}^{h_e=-1}-\mathcal{N}^{h_e=0})/\mathcal{N}^{h_e=0}$ and $\mathcal{N}=\mathrm{d}N_{\gamma}/\mathrm{d}\varepsilon_{\gamma}$. Other laser and electron-beam parameters are given in the text.}\label{fig3}
	\end{center}
\end{figure}

\begin{figure}[t] 
	\begin{center}
		\includegraphics[width=\linewidth]{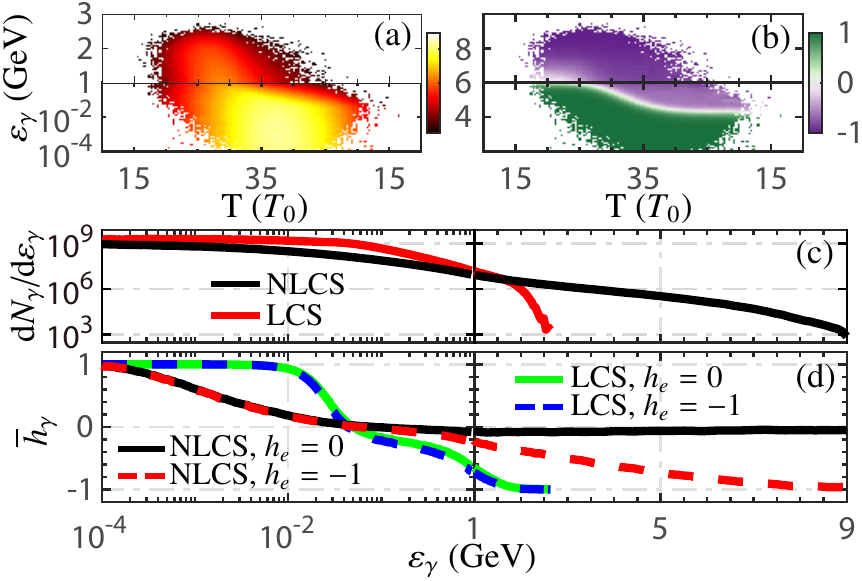}
		\caption{
			Distributions of $\gamma$ photons predicted by LCS for SNP electrons ($h_e=0$) with (a) the yields log$_{10} \frac{\mathrm{d}N_\gamma}{\mathrm{dTd}\varepsilon_{\gamma}}$ and (b) average helicity $\overline{h}_\gamma$, respectively. (c) and (d) Comparisons of energy spectra d$N_\gamma$/d$\varepsilon_{\gamma}$ and $\overline{h}_\gamma$ predicted by LCS and NLCS, respectively. The laser and electron-beam parameters are the same as those in Figs.\ref{fig3}(d)-\ref{fig3}(f).}\label{fig4}
	\end{center}
\end{figure}

In this Letter, the manipulation of $\gamma$-ray polarization in CS employing the S(N)P ultrarelativistic electron beam  is investigated. We find that the contradiction lies in the simultaneous achievement of high-brilliance and high-polarization of $\gamma$ rays by increasing laser intensity, since the polarization is predominately contributed by the electron (laser photon) spin via multi-photon (single-photon) absorption channel (see Figs.\ref{fig1} and \ref{fig2}).  And, SNP electrons in high-intensity laser pulse can also generate high-brilliance high-polarization $\gamma$ photons (see Fig.\ref{fig3}). The transfer mechanism of SAM is also valid in the generation of vortex $\gamma$ photons due to the same radiation dynamics as the plane-wave $\gamma$ photons \cite{Ababekri2022}.  Moreover, the polarization of $\gamma$ photons radiated by the electrons with different initial spin states proceeds in different ways in LCS and strongly NLCS respectively, which can give a clear evidence of the nonlinear effects in CS with moderate-intensity laser facilities (see Fig.\ref{fig4}).

When electrons scatter with a laser pulse, they may absorb single or multiple low-energy laser photons and then emit a high-energy $\gamma$ photon via CS. The polarization-dependent transition rate summing over the final electron spin in a circularly polarized (CP) monochromatic field is given by \cite{CAIN242,Ivanov2004}
\begin{equation}
\frac{\mathrm{d}W}{\mathrm{d}t} = W_0\sum_{n=1}^{\infty} \int_{0}^{\delta_n} \mathrm{d} \delta[F_{1n}+h_L h_e F_{2n}+h_\gamma (h_L F_{3n}+h_e F_{4n})], \label{eqn1}
\end{equation}  
where, $W_0$ = $\alpha m_e^2 a_0^2/(8\varepsilon_\mathrm{eff})$, $a_0 = \left| e \right| E / (m_e \omega_L)$ is the laser invariant intensity, $\varepsilon_\mathrm{eff} = \varepsilon_e+a_0^2 \omega_L/\Lambda$ the effective energy of initial electron in the laser field, $\delta=(k_L \cdot k_\gamma)/(k_L \cdot p) \approx \varepsilon_\gamma/\varepsilon_e$ the energy ratio parameter, $\delta_n = n \Lambda / (1+a_0^2+n \Lambda)$ the cutoff-energy fraction of emitted photon absorbing $n$ laser photons, i.e., $n$-th Compton edge \cite{Ivanov2004}, $\Lambda=2(k_L \cdot p)/m_e^2$ the invariant variable, $p$, $k_L$ and $k_\gamma$ the four-momenta of the initial electron, laser photon and emitted photon, respectively, $\alpha$ the fine structure constant, $e$ and $m_e$ the charge and rest mass of the electron, and $\omega_L$ and $E$ the frequency and amplitude of the laser field, respectively. Relativistic units with $c=\hbar=1$ are used throughout. $F_{kn}$ ($k = 1,2,3,4$) in Eq.~(\ref{eqn1}) are detailed in \cite{Supplemental}. $h_L$, $h_e$ and $h_\gamma$ are helicities of the laser, initial electron, and emitted photon, respectively. The helicity of emitted $\gamma$ photon is determined by \cite{CAIN242,Ivanov2004}
\begin{equation}
h_\gamma = \frac{h_LF_3+h_eF_4}{F_1+h_Lh_eF_2}, \label{eqn2} 
\end{equation}
where, $F_k = \sum_{n=1}^{\infty} F_{kn}$ ($k=1,2,3,4$, respectively). The polarization of $\gamma$ photon hardly changes during the subsequent propagation \cite{CAIN242,Ivanov2005,Supplemental}.

The transfer mechanism of SAM in CS from linear to strongly nonlinear processes is illustrated in Fig.\ref{fig1}.  For $a_0 \lesssim O(0.1)$, there is a distinct edge at the end of the first harmonic followed by smaller probabilities of higher harmonics [see Figs.\ref{fig3}(a) and .\ref{fig3}(b)]. For instance, for $a_0=0.1$, $\frac{\mathrm{d}^2W_{rad}}{\mathrm{d}t\mathrm{d}\delta} \approx 10^{-5.31}$ at $\delta \approx 0.192$ is two orders of magnitude smaller than  $\frac{\mathrm{d}^2W_{rad}}{\mathrm{d}t\mathrm{d}\delta} \approx 10^{-3.01}$ at $\delta =\delta_1 \approx 0.190$, and for larger $\delta$, $\frac{\mathrm{d}^2W_{rad}}{\mathrm{d}t\mathrm{d}\delta}$ is smaller, where $W_{rad}$ is the radiation probability after summing over the emitted photon helicity and  averaging over the initial electron spin with $\frac{\mathrm{d}^2W_{rad}}{\mathrm{d}t\mathrm{d}\delta} = 2W_0 \sum_{n=1}^{\infty} F_{1n}$, and $\delta_1$ is the first Compton edge of the emitted photon. Therefore, for $a_0 \lesssim O(0.1)$, the electron absorbs almost only one laser photon and radiates a $\gamma$ photon with $\delta \leq \delta_1$, i.e., the scattering process is LCS. For $a_0 \sim O(1)$, the process of absorbing dozens of laser photons appears with $\frac{\mathrm{d}^2W_{rad}}{\mathrm{d}t\mathrm{d}\delta}\approx 10^{-2.99}$ of the average number of absorbed laser photons $\overline{n}=10$ (corresponding to $\delta \approx 0.4$), termed as weakly NLCS. As $a_0$ increases to $O(10)$, due to $W_{rad} \propto a_0^2$, the electron will have a greater probability of absorbing thousands of laser photons to emit a higher-energy $\gamma$ photon with $\frac{\mathrm{d}^2W_{rad}}{\mathrm{d}t\mathrm{d}\delta}\approx 10^{-1.22}$ of $\overline{n}=1000$ (corresponding to $\delta \approx 0.55$), described as strongly NLCS. Therefore, electrons will radiate more brilliant $\gamma$ rays in higher-intensity laser field. Due to $\delta_n \propto 1/a_0^2$ for a given $n$, the harmonic cutoffs decrease as $a_0$ increases, e.g., $\delta_1 \approx 0.190$ for $a_0=0.1$ while $\delta_1 \approx 0.0023$ for $a_0=10$. Besides, for a certain $a_0$ the gaps of harmonic cutoffs $\Delta \delta_n = \delta_n-\delta_{n-1} \propto 1/n^2$ decrease with absorbing more laser photons [see examples of four lines corresponding to the first four cutoffs $\delta_{n=1,2,3,4}$ in Fig.~\ref{fig1}(a)]. Therefore, the harmonic structure is clearly visible in linear and weakly nonlinear CS spectra, however, which becomes smoother in strongly NLCS. Importantly, the impact of electron spin on the transition rate weakens as the laser intensity increases. For instances, for $a_0=0.1$, the radiation probability of the longitudinally spin-polarized (LSP) electrons (satisfying $h_L h_e=-1$) at the first edge can be increased by $\Delta W_i=(W_i^{\mathrm{LSP}}-W_i^{\mathrm{SNP}})/W_i^{\mathrm{SNP}} = -\frac{F_2}{F_1} \approx 20\%$, yet, for $a_0=10$, $\Delta W_i \sim 0$ [see Fig.\ref{fig1}(c)], where $W_i^{\mathrm{LSP}}$ ($W_i^{\mathrm{SNP}}$) is the radiation probability of LSP (SNP) electrons after summing over the emitted photon polarization in Eq.(\ref{eqn1}).

The electron spin plays an increasingly significant role in the transfer of SAM with stronger nonlinear effects. The entanglement term $\frac{F_2}{F_1}$ of the laser helicity and electron spin in the photon helicity Eq.(\ref{eqn2}) mainly operates at the end of the first few harmonics and decreases as $a_0$ increases. While $\frac{F_2}{F_1}$ can reach to $-0.68$ at $\delta=0.7$ for $a_0=0.1$, it is irrelevance due to the relatively low radiation probability with $\frac{\mathrm{d}^2W_{rad}}{\mathrm{d}t\mathrm{d}\delta} \approx 10^{-25}$  [see Figs.\ref{fig1}(a) and \ref{fig1}(c)]. Therefore, the helicity of emitted $\gamma$ photon is mainly related to the independent laser helicity term $F_3$ and electron spin term $F_4$. For emitted photons with $\delta \lesssim \delta_1$ at any $a_0$, the laser helicity contribution $\frac{\left| F_3 \right|}{\left| F_3 \right|+F_4} \approx 1$ and correspondingly the electron spin contribution $\frac{F_4}{\left| F_3 \right|+F_4} = 1-\frac{\left| F_3 \right|}{\left| F_3 \right|+F_4} \approx 0$, i.e., regardless of whether the scattering process is linear or not, the average helicity of $\gamma$ photons via the single-photon channel is almost completely determined by the laser [see Fig.\ref{fig1}(d)].  When electrons simultaneously absorb dozens of laser photons, higher-energy $\gamma$ photons are emitted, where the laser helicity contribution gradually decreases and the electron spin comes into play, i.e., the SAM transfer enters into a competitive stage of the laser and electron for controlling the $\gamma$-photon polarization [see the black-dotted line in Fig.\ref{fig1}(d), which almost corresponds to $\overline{n}=10$ in Fig.\ref{fig1}(b)]. As the number of absorbed laser photons continues to increase, the electron spin plays a dominant role in the SAM transfer, for instance, as $\overline{n}$ up to 1000, $\frac{\left| F_3 \right|}{\left| F_3 \right|+F_4}\approx 0.05$ and $\frac{F_4}{\left| F_3 \right|+F_4}\approx 0.95$ at $\delta=0.55$ for $a_0=10$.
 
The transfer mechanism of SAM points a direction for manipulating the polarization of $\gamma$ rays. For LSP ($h_e=-1$) electrons scattered with the CP ($h_L=1$) laser, in LCS ($a_0 \lesssim 0.1$), the emitted photons are almost completely radiated by the single-photon absorption channel and helicities are mainly contributed by the laser [see Fig.\ref{fig2}(a)]. In the low-energy parts of the first harmonic, $\gamma$ photons are radiated in the forward scattering with average helicity $\overline{h}_\gamma \simeq h_L = 1$, while near the edge $\overline{h}_\gamma \simeq -h_L = -1$ via the backward scattering \cite{Babusci1995}. As $a_0$ increases, the interaction process transitions into the NLCS regime, due to $\delta_1 \propto 1/a_0^2$, the energy regions of $\overline{h}_\gamma \sim 1$ of radiated photons via the single-photon channel tend toward lower-energy areas with smaller $\delta$. And for a certain $a_0 \gtrsim 1$, $\gamma$ photons with high energy ($\delta \gtrsim \delta_{100}$) obtain increasingly high $\overline{h}_\gamma$ transferred by the electron spin. For the case of $h_e=1$, the helicity behavior is similar to that of $h_e=-1$ but slightly different especially at the end of the first few harmonics with $a_0 \lesssim 1$ due to the entanglement term of the laser helicity and electron spin \cite{Supplemental}. If electrons are SNP ($h_e=0$), $\overline{h}_\gamma$ of $\gamma$ photons is only provided by the laser helicity. Therefore, $\overline{h}_\gamma$ of high-energy $\gamma$ photons via multi-photon absorption channels ($\overline{n} \gtrsim 10$) gradually decreases as $a_0$ increases, for instance, at $\delta=0.3$, $\overline{h}_\gamma \approx -0.40$ and $-$0.03 for $a_0=1$ and 10, respectively [see Fig.\ref{fig2}(b)].       

Above analytical discussions are based on the single-photon emission of electrons with $\varepsilon_e =10$ GeV in a monochromatic plane-wave field. However, the transfer mechanism also holds true for other electron energies. The slight difference lies in the positions of harmonic edges and average number of absorbed laser photons \cite{Supplemental}. For linearly polarized (LP) $\gamma$ photons generated in the LP laser, the competition between the laser and electron for controlling the $\gamma$-photon polarization also exists \cite{Supplemental}. If electrons are SP, one can increase the laser intensity  to generate high-brilliance high-polarization $\gamma$ rays. However, for SNP electrons, due to the energy regions $\delta \lesssim \delta_1 \propto \varepsilon_{e}$ for a certain $a_0$, apart from the high-intensity laser, higher-energy electrons are also required to produce the similar high-quality $\gamma$ rays, where the polarization is mainly contributed by the laser via the single-photon absorption channel. Importantly, when scattering process satisfies the rotational symmetry around the collision axis, e.g., in the case of CP laser, the angular momentum conservation holds and one could expect the generation of vortex $\gamma$ photons with intrinsic orbital angular momentum (OAM) via multi-photon absorption channels. In the monochromatic CP laser, $\gamma$ photons corresponding to the $n$-th harmonic carry OAM $l_\gamma=n h_L + h_e - h^\prime_{e}-h_\gamma$ with the final electron helicity $h^\prime_{e}$. Since the radiation dynamics of the vortex $\gamma$ photons are the same as those of the plane-wave $\gamma$ photons \cite{Ababekri2022}, the transfer mechanism of SAM discussed above is still valid. However, the electrons are required to possess coherence in the transverse plane with axial symmetry in order to attain vortex $\gamma$ photons.   

The manipulation mechanism paves a clear path for high-brilliance high-polarization $\gamma$-ray sources. And the generation of CP $\gamma$ rays via different CS mechanisms in realistic laser pulse is detailed in Fig.\ref{fig3}. The focused Gaussian left-hand CP laser pulse ($h_L=1$) is employed with peak intensity $a_{0,peak}=1$, wavelength $\lambda_0 = 0.8 ~ \mu$m, pulse duration $\tau=10~T_0$ with the period $T_0$, and focal radius $w_0=5 ~ \mu$m. For the head-on colliding cylindrical electron beam, the initial kinetic energy $\overline{\varepsilon}_e=10$ GeV, energy spread $\Delta \overline{\varepsilon}_e / \overline{\varepsilon}_e=6\%$, radius $r_e = 2 ~ \mu$m with transversely Gaussian profile, longitudinal length $l_e = 3 ~ \mu$m, polar angle $\theta_e=180^\circ$ with angular divergence $\Delta \theta_e=0.2$ mrad, and the electron number $N_e=5\times10^6$. Such electron bunches can be obtained via laser wakefield acceleration \cite{Leemans2014,Xia2016,Gonsalves2019,LiF2022}. In the laser pulse, the intensity sensed by electrons changes in real-time and is strongest at about T $=35~T_0$ where copious $\gamma$ rays are radiated [see Fig.\ref{fig3}(a)].  For SNP electrons, the helicities of $\gamma$ photons come from the laser. In the front of the laser pulse with T $\lesssim 23~T_0$ and $a_0 \lesssim 0.1$, electrons only radiate a small amount of high polarized $\gamma$ photons due to the low radiation probability of LCS regime [see Figs.\ref{fig3}(a) and \ref{fig3}(b)]. And the polarization directions of $\gamma$ photons with $\varepsilon_\gamma \gtrsim 1$ GeV ($\delta \gtrsim 0.1$) and $\varepsilon_\gamma < 1$ GeV ($\delta < 0.1$) are opposite [see Figs.\ref{fig2}(b) and \ref{fig3}(b)]. For $23~T_0 \lesssim$ T $\lesssim 35~T_0$, $a_0$ gradually increases to 1 and since $\delta_1 \propto 1/a_0^2$ the energy regions of $\overline{h}_\gamma \simeq -h_L =-1$ decrease from $\varepsilon_\gamma \lesssim 2$ GeV to $\lesssim 1$ GeV. And there are high-energy $\gamma$ photons radiated by electrons absorbing dozens of laser photons and $\overline{h}_\gamma$ of GeV $\gamma$ photons decreases from $\sim -1$ to $\sim -0.43$ due to the decline of the transfer efficiency of laser helicity. As T continues to increase, $a_0$ begins to decreases and the trends of radiated spectra and helicities of $\gamma$ photons are basically symmetrical with those of T $\lesssim 35~T_0$. Therefore, the final $\overline{h}_\gamma$ of $\gamma$ photons is the comprehensive result due to the laser pulse effect, and the clearly first few harmonics are smoothed out and $\overline{h}_\gamma$ of $\gamma$ photons with $\varepsilon_{\gamma} \gtrsim 2$ GeV is about $-0.53$ which is higher than $\overline{h}_\gamma \approx -0.43$ of the plane wave with $a_0=1$ [see Fig.\ref{fig2}(b) and the green-solid line in Fig.\ref{fig3}(h)]. For LSP electrons \cite{Wen2019,Nie2021,Sun2022}, the helicities of $\gamma$ photons contributed by the CP laser are almost identical to the case of SNP electrons for all energy regions radiated at the both ends of the laser pulse with T $\lesssim 23~T_0$ and T $\gtrsim 49~T_0$ and $\varepsilon_\gamma \lesssim 1$ GeV radiated at the middle of the laser pulse with $23~T_0 \lesssim$ T $\lesssim 49~T_0$. However, at the middle of the laser pulse the $\gamma$ photons with $\varepsilon_\gamma \gtrsim 1$ GeV receive helicities from the scattering electrons and $\overline{h}_\gamma$ linearly falls (rises) as $\varepsilon_\gamma$ increases for $h_e=-1$ ($h_e=1$) [see Fig.\ref{fig3}(c), the blue-dash-dotted line in Fig.\ref{fig3}(h), and \cite{Supplemental}]. The relative deviation of energy spectra $\Delta_{\gamma}$ can reach to 30\% (-30\%) for $h_e=-1$ ($h_e=1$) [see the red dash-dotted line in Fig.\ref{fig3}(g)]. 

Usually, one can increase the laser intensity to obtain high-brilliance $\gamma$ rays. For instance, the $\gamma$-photon yields in CS with  $a_{0,peak}=10$ are two to three orders of magnitude higher than those of the $a_{0,peak}=1$ case [see Figs.\ref{fig3}(d) and \ref{fig3}(g)]. And the yield and helicity distributions are not completely symmetric due to the radiation reaction. Importantly, SNP electrons can also generate high-brilliance high-polarization $\gamma$ rays with the energy scale of keV to MeV [see Fig.\ref{fig3}(e) and the black-solid line in Fig.\ref{fig3}(h)]. Similar to the $a_{0,peak}=1$ case, one gets $\left|\overline{h}_\gamma\right| \approx h_L =1$ for $\gamma$ photons via linear and weakly nonlinear CS at the front and rear of the laser pulse. During $23~T_0 \lesssim $ T $\lesssim 49~T_0$, the scattering sequentially undergoes weakly nonlinear, strongly nonlinear, and again weakly nonlinear processes. At about 35 $T_0$, due to $\delta_1 \propto 1/a_0^2$ the energy regions of $\overline{h}_\gamma \simeq h_L= 1$ drop below 100 keV, which is much smaller than that in the plane wave case ($\varepsilon_\gamma = \delta \varepsilon_e \lesssim$ 10 MeV) in Fig.~\ref{fig2}, due to radiation reaction. Meanwhile, electrons experience the strongest field and have larger probabilities of absorbing thousands of laser photons to radiate higher-energy $\gamma$ photons; see Fig.\ref{fig3}(g). For $\varepsilon_\gamma \gtrsim 1$ GeV, $\overline{h}_\gamma$ decreases from $-$0.28 at T $\approx 23~T_0$ to $-$0.05 at T $\approx 35~T_0$ since the transfer efficiency of laser helicity decreases as $a_0$ increases. The corresponding brilliances $\mathcal{B}$ (average helicities $\overline{h}_\gamma$) are $1.49 \times 10^{20}$ (0.98), $1.14 \times 10^{21}$ (0.59) and $5.23 \times 10^{21}$ (0.16) photons/(s $\cdot$ mm$^2$ $\cdot$ mrad$^2$ $\cdot$ 0.1\%BW) for $\varepsilon_\gamma=100$ keV, 1 MeV, and 10 MeV, respectively. For LSP electrons, the $\gamma$ photons spectra are almost identical to the case of $h_e=0$ with $\Delta_{\gamma} \approx 5\%$ at $\varepsilon_\gamma=9$ GeV [see the red-solid line in Fig.\ref{fig3}(g)]. However, electrons transfer SAM to high-energy $\gamma$ photons with $\varepsilon_{\gamma} \gtrsim 1$ GeV via strongly NLCS and $\overline{h}_\gamma$ is linearly falling (rising) as $\varepsilon_{\gamma}$ in the case of $h_e=-1$ ($h_e=1$) and reaches $\simeq -1$ (1) at $\varepsilon_{\gamma}=9$ GeV [see Fig.\ref{fig3}(f), the red-dash-dotted line in Fig.\ref{fig3}(h), and \cite{Supplemental}]. Note that with $a_{0,peak}=10$, the locally constant field approximation method commonly used to simulate strongly NLCS performs poorly not only in radiation spectra but also in photon polarization \cite{Ritus1985,Baier1989,Harvey2015,Dinu2016,Piazza2018,Piazza2019,Ilderton2019a,Ilderton2019b,Heinzl2020,Blackburn2021,Supplemental}.

The different helicity behaviors of $\gamma$ rays in linear and nonlinear CS regimes also represent a clear signal for testing the nonlinear effects of CS (see Fig.\ref{fig4}). In LCS, the first cutoff $\delta_1 = \Lambda/(1+\Lambda) \approx 4\varepsilon_{e} \varepsilon_L/(m_e^2+4\varepsilon_{e} \varepsilon_L)$ and higher harmonics ($n > 1$) disappear. Therefore, there is a distinct edge at $\varepsilon_{\gamma,edge}=\delta_1 \varepsilon_{e}=1.9$ GeV, which is smaller than 2.5 GeV of the numerical result due to the initial electron energy spread [see the red-solid line in Fig.\ref{fig4}(c)]. Furthermore, the helicities of emitted photons are almost completely derived  from the laser, and due to the opposite scattering direction, $\overline{h}_\gamma$ at both ends of the spectra are anti-parallel. And because of the radiation reaction, $\varepsilon_{\gamma,edge}$ and the turning points from $\overline{h}_\gamma \approx 1$ to $-1$ decrease as T increases [see Figs.\ref{fig4}(a) and \ref{fig4}(b)]. As a result, for 10 MeV $\lesssim \varepsilon_\gamma \lesssim 1$ GeV, $\overline{h}_\gamma$ is averaged to a smaller polarization degree [see the green-solid line in Fig.\ref{fig4}(d)]. Therefore, the strongly nonlinear effects not only broaden the spectra but also change the helicity distributions of radiated $\gamma$ photons compared with the LCS process [see Figs.\ref{fig4}(c) and \ref{fig4}(d)]. More significantly, if electrons are LSP, the nonlinear signals will be more sensitive. For instance, the electron spin hardly affects the $\gamma$-photon helicity in the LCS process, while, in strongly NLCS, LSP electrons will absorb abundant laser photons and transfer SAM to high-energy $\gamma$ photons [see Fig.\ref{fig4}(d)].
  
In conclusion,  we have investigated the transfer mechanism of SAM in CS from linear, through weakly into strongly nonlinear regimes to achieve the simultaneous manipulation of high-brilliance and high-polarization of $\gamma$ rays, paving a way for high-quality polarized $\gamma$-beam sources. Moreover, detecting SAM of particles can help us to observe the nonlinear effects of strong field QED processes with currently feasible laser facilities.\\

{\emph{Acknowledgment:} The work is supported by the National Natural Science Foundation of China (Grants No. 12022506, No. U2267204, No. U2241281), the Foundation of Science and Technology on Plasma Physics Laboratory (No. JCKYS2021212008), the Shaanxi Fundamental Science Research Project for Mathematics and Physics (Grant No. 22JSY014), and Open Foundation of Key Laboratory of High Power Laser and Physics, Chinese Academy of Sciences (SGKF202101).}

\bibliography{library}
\end{document}